\documentclass[aps,prl,twocolumn,groupedaddress,showpacs]{revtex4-1}

\usepackage{graphicx}
\usepackage[geometry]{ifsym}
 \usepackage{epsfig}
 \usepackage{color}
 \usepackage{soul}
 \usepackage{amsmath}

\begin{document}

\noindent {\bf Lau, Berciu and Sawatzky reply:} In the preceding
Comment\cite{leecomment}, Lee and Lee bring to attention their
interesting variational calculation for the single band Hubbard model
further reduced to a $t$-$t'$-$t''$-$J$ model \cite{lee2003}. Its main
result was to reveal new  low-energy one-hole states called
spin-bags (SB),  possibly forming a continuum. SBs consist of
a quasiparticle (QP) plus a spin-wave excited in the AFM
background, and are found to cross below the QP band in  
some regions of the Brillouin zone (BZ). Based on this, Lee and Lee
claim that the SBs  explain, within a one-band model, the
spin-${3\over 2}$ polaron that we recently found in a 
three-band model \cite{lau2011}. They conclude that our claim that
this model reveals physics that cannot be
described within one-band models, {\em i.e.} in the framework of
Zhang-Rice 
singlets (ZRS),  is not justified.

While superficial similarities exist between the SB and the
spin-${3\over 2}$ polaron, we disagree that they describe the same
physics, on several grounds: 

(i) In the Supplemental Material of Ref. \cite{lau2011}, we
ruled out the possibility that the spin-${3\over 2}$
polaron is a spin-${1\over 2}$ polaron plus a free magnon,
because its band lies below the continuum describing such states. It
can be roughly thought of as a bound-state of a spin-${1\over 2}$
polaron and a magnon, with a very distinct local spin structure around
the charge. The existence of such bound states, which might be a
better analog of our spin-${3\over 2}$ polaron,
is not analyzed for the one-band model,  in Refs. \cite{leecomment,lee2003}; 

(ii) As shown in our Fig.~2, the spin$-{3\over2}$ polaron's band has
significant dispersion, comparable to that of the spin$-{1\over2}$
polaron \cite{lau2011}. In contrast, the low-energy edge of the SB
continuum is rather flat throughout the BZ, see Fig. 1(b) of
Ref. \cite{lee2003}. This striking difference in their spectra is
likely an indication of a very different nature of the two types of
low-energy states. There is currently no evidence that the two models
have comparable dispersion for spin excitations, regardless of their
nature.
%Moreover, without a proper infinite-lattice solution to the three-band model, there
% no reason to further postulate the two models to have comparable dispersion
%for spin excitations.

%Moreover, note that even if bound states were to be found in the 1BM
%of Ref. \cite{lee2003}, they would have to lie above the SB continuum. As
%such, they could not possibly disperse on a scale comparable to that of their QP
%band, either; 

(iii) While the spin-${3\over 2}$ polaron band crosses below the
spin-${1\over 2}$ polaron band in certain regions, just like the SB
continuum is below the QP band in certain regions of the BZ, a careful
comparison shows yet more differences. In our model, this happens in
two separate regions, centered at $(0,0)$ and $(\pi,\pi)$. In the
variational solution for the one-band model, this happens in one
larger region centered at $k=(\pi,\pi)$ which, coincidentally, is the
AFM order vector. The difference is most clearly visible along the
$(\pi,0)-(0,\pi)$ cut, where we find no crossing whereas the
variational calculation predicts the QP as the low-energy state only
near $({\pi\over 2},{\pi\over 2})$.  If bound-states were found in the
one-band model, the comparison would be worse since this would further
increase the crossing region.

Such differences result in very different physics, {\em eg.}  at the
nodal point. While the vanishing quasiparticle weight at $(0,\pi)$ is
explained as being due to the SB state in the one-band model, we find
$Z=0$ here because of the orthogonal reflection parity between the
lowest electron-removal state and the lowest spin-${1\over2}$
eigenstate \cite{2h}.

A second point raised in the Comment is that if a ZR-like state is
built from a superposition of configurations like that of Fig. 3a, AFM
correlations on the e and d bonds are similar to those calculated in
Ref. \cite{lee2003}. This is taken as proof that the
spin-${1\over2}$ polaron is similar to the ZR-based QP state, as well.
 First, Fig. 3a is for a state of momentum $({\pi\over2},
{\pi\over2})$, so naive ${\pi\over 2}$ rotations  lead to a state
with an ill defined momentum. In fact, even though these bonds are
related by the exact $\hat{P}_{x \leftrightarrow y}$ symmetry of our
Hamiltonian, the quoted values are not equal; this is wrong.  In any
case, the fact that bonds rather far from the hole show robust AFM
correlations is hardly surprising. The key observation in our model is
the strong FM correlation between the spins neighboring the hole,
which points to the three-spin polaron (3SP) as the proper framework
to understand the spin-${1\over 2}$ polaron and the inner core of the
spin-${3\over 2}$ polaron. Since the 3SP can be written as the sum of
singlets between the hole and each of its neighboring spins
\cite{lau2011}, it does have a finite overlap with a ZR state~\cite{ding}. Its
additional degrees of freedom, however, allow it to describe
correlations beyond those possible in a ZR-based model. This invalidates
the Comment's claim that a low-energy non-bonding state  is the only
signature of breakdown for a one-band model. It is very 
important, in this context,  to also point out that the model
used in Ref.~\cite{lee2003} is a further simplification of the ZR scenario --
the O states are no longer present and a discussion of the  spin
correlations around an O hole becomes meaningless.

%This is confirmed in Ref. \cite{ding}, where the exact ground-state for a
%16-site model is shown to have 97\% overlap with a 3SP-type state, and
%only 67\% overlap with a ZR-like state.

%Well actually as per
%the comment above which I would like to see included there is
%no way in a single band model that one can look at this even never mind telling us if a ferromagnnetic correlation exists. 

For all these reason we remain convinced that both the spin-$\frac{1}{2}$ and spin-$\frac{3}{2}$ polarons in our 3 band model are quite different objects from the QP and SB obtained from a single band description However, caution is necessary since comparisons
between a variational solution based on mean-field and our exact
diagonalization (ED) for a finite cluster may be misleading. If ED
results for a one-band model revealed similar low-energy spin-${3\over
  2}$ states, and FM correlations between the spins sandwiching the
hole, our position would have to be reconsidered. To our knowledge,
the former is not the case and the latter is not possible for a
one-band model.

\vspace{3mm}

\noindent Bayo Lau, Mona Berciu and George A. Sawatzky\\
\small{Department of Physics and Astronomy\\
University of British Columbia\\
Vancouver, British Columbia, Canada V6T 1Z1}

% Create the reference section using BibTeX: \bibliography{basename of

\begin{thebibliography}{99}

\bibitem{leecomment} W.-C. Lee and T. K. Lee, preceding Comment (arXiv:1108.5413v1).
\bibitem{lee2003} W.-C. Lee, T. K. Lee, C.-M. Ho, and
  P. W. Leung. Phys. Rev. Lett. {\bf 91}, 057001 (2003).

\bibitem{lau2011} B. Lau, M. Berciu, and
  G. A. Sawatzky. Phys. Rev. Lett. {\bf 106}, 036401 (2011).

\bibitem{2h} B. Lau, M. Berciu, and G. A. Sawatzky,  arXiv:1107.4141 

\bibitem{ding} See, for example, Ref. [18] in our Letter.

%H. Q. Ding, G. H. Lang, and W. A. Goddard, Phys. Rev. B  {\bf 46}, 14317 (1992)

%\bibitem{ding} For example, H. Q. Ding, G. H. Lang, and W. A. Goddard, Phys. Rev. B
  %{\bf 46}, 14317 (1992) shows that the exact ground-state for a
%similar model to have 97\% overlap with a 3SP-type state, but
%only 67\% overlap with a ZR-like state.

\end{thebibliography}
%.bib file}

\end{document}